\documentclass[10pt,journal]{IEEEtran}
\IEEEoverridecommandlockouts
\usepackage{amsfonts}
\usepackage{bbm}
\usepackage{amssymb}
\usepackage{array, makecell}
\usepackage{mathtools}
\usepackage{graphicx}
\usepackage{subfigure}
\usepackage{enumerate}
\usepackage{amsmath}
\usepackage{color}
\usepackage{amsthm}
\usepackage[colorlinks,linkcolor=black,anchorcolor=black,citecolor=black]{hyperref}
\usepackage{algorithm}
\usepackage{algpseudocode}
\usepackage{stfloats}
\usepackage{multirow}
\theoremstyle{definition}
\usepackage{color}
\usepackage{url}
\usepackage{bm}
\usepackage{adjustbox}

\allowdisplaybreaks[4]
\newcommand\reg{\textsc{Regret}}

\ifCLASSOPTIONcompsoc
  \usepackage[nocompress]{cite}
\else
  \usepackage{cite}
\fi

\usepackage[bottom=1.18in,left=0.65in,right=0.6in,top=0.67in]{geometry}
    
\begin{document}
\newtheorem{theorem}{Theorem}
\newtheorem{lemma}{Lemma}
\newtheorem{assumption}{Assumption}
\newtheorem{definition}{Definition}

\title{Quantum Entanglement Path Selection and Qubit Allocation via Adversarial Group Neural Bandits\\
\thanks{* The first two authors contributed equally to this work.}
}



\author{Yin Huang*, Lei Wang*, 
    Jie Xu,~\IEEEmembership{Senior Member,~IEEE}
\thanks{Yin Huang, Lei Wang, and Jie Xu are with the Department of Electrical and Computer Engineering, University of Florida, Gainesville,
FL 32608, USA. Email: \{yin.huang, leiwang1, jie.xu\}@ufl.edu.
}
}

\maketitle

\begin{abstract}

Quantum Data Networks (QDNs) have emerged as a promising framework in the field of information processing and transmission, harnessing the principles of quantum mechanics. QDNs utilize a quantum teleportation technique through long-distance entanglement connections, encoding data information in quantum bits (qubits). Despite being a cornerstone in various quantum applications, quantum entanglement encounters challenges in establishing connections over extended distances due to probabilistic processes influenced by factors like optical fiber losses. The creation of long-distance entanglement connections between quantum computers involves multiple entanglement links and entanglement swapping techniques through successive quantum nodes, including quantum computers and quantum repeaters, necessitating optimal path selection and qubit allocation. Current research predominantly assumes known success rates of entanglement links between neighboring quantum nodes and overlooks potential network attackers. This paper addresses the online challenge of optimal path selection and qubit allocation, aiming to learn the best strategy for achieving the highest success rate of entanglement connections between two chosen quantum computers without prior knowledge of the success rate and in the presence of a QDN attacker. The proposed approach is based on multi-armed bandits, specifically adversarial group neural bandits, which treat each path as a group and view qubit allocation as arm selection. Our contributions encompass formulating an online adversarial optimization problem, introducing the EXPNeuralUCB bandits algorithm with theoretical performance guarantees, and conducting comprehensive simulations to showcase its superiority over established advanced algorithms.
\end{abstract}

\section{Introduction}\label{s1}

\IEEEPARstart{I}{n} recent times, Quantum Data Networks (QDNs) have emerged as a transformative approach, with the potential to reshape information processing and transmission through the evolution of distributed quantum computing \cite{qiao2022quantum}. Classical networks have inherent limitations when it comes to ensuring data security during transmission and handling data-intensive processing tasks. QDNs, built upon the principles of quantum mechanics, have the potential to overcome these limitations by leveraging the unique properties of quantum systems, paving the way for unparalleled levels of security, communication efficiency, and computational power \cite{steane1998quantum}. Previous quantum key distribution technique, that aims to generate quantum bits (qubits) and use them to deliver cryptographic keys, can regenerate and retransmit qubits if available \cite{mehic2020quantum}, while QDNs encode the data information in the data qubits and utilize quantum teleportation technique to avoid the retransmission issue due to no-cloning theorem \cite{wootters1982single}.

Quantum entanglement stands as a cornerstone in various quantum techniques and applications. Consider two physically adjacent quantum computers, Alice and Bob, where entangled qubit pairs are generated on one end, say Alice. One of these entangled qubits is retained while the other is sent to Bob, via a physical fiber-optic channel, establishing an entanglement link. However, this procedure encounters challenges due to the inherent loss in the optical fiber, resulting in a success rate well below one \cite{stephenson2020high}. Creating an entanglement link is probabilistic and uncertain, relying on the presence of quantum channels between the parties and qubits at both ends \cite{mattle1996dense}. Increasing the number of qubits on both ends and utilizing additional available channels can improve the success rate of the simultaneous attempts made. Nevertheless, quantum channels are limited, and each quantum node has constrained quantum memory for storing qubits, further complicating the process. 

\begin{figure}
    \centering
    \includegraphics[width=0.4\textwidth]{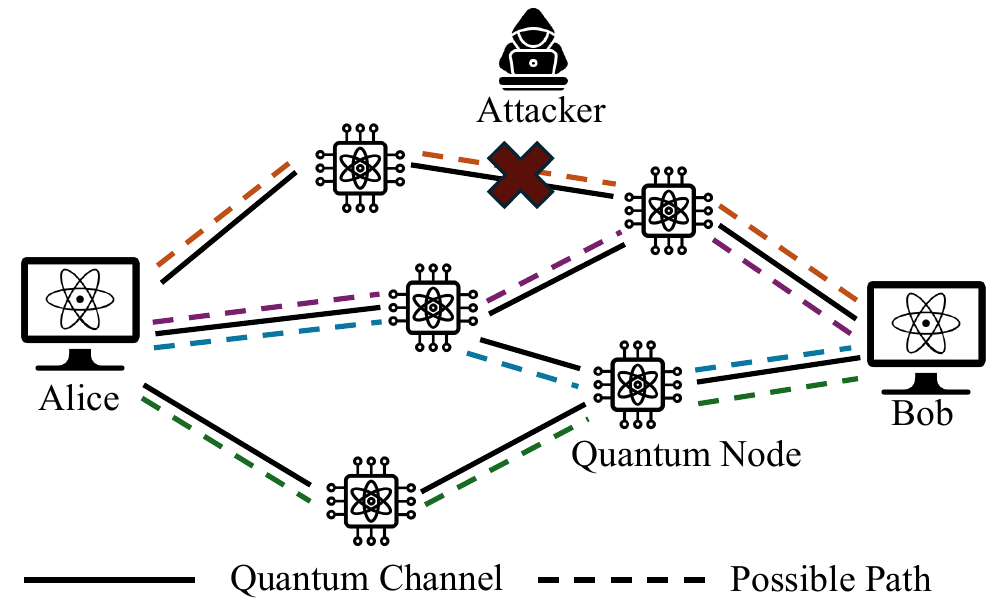}
    \vspace{0pt}
    \caption{Quantum Data Network. There exist four possible paths between Alice and Bob and one attacker aims to disrupt one of them. Note that different possible paths can have different success rates of establishing entanglement connections}
    \label{fig:QDN}
    \vspace{-5pt}
\end{figure}

As depicted in Figure~\ref{fig:QDN}, within QDNs, direct channel connections among quantum computers are often absent. Instead, they are linked through various quantum nodes, known as quantum repeaters. Notably, entanglement links can only be forged between neighboring quantum nodes. To establish a long-range entanglement connection between Alice and Bob, a path must be discovered between them, establishing entanglement links for each successive quantum node along this path and entanglement swapping operations are employed repeatedly at each quantum node along the path \cite{jennewein2001experimental}. The success rate of establishing such a long-distance entanglement connection is related both to the selected path, such as the length and number of hops of the path and the success rate of each entanglement link on the path and to the allocation of qubits on each quantum node because of the limited quantum memory for storing qubits. This means that more qubits allocated by a quantum node to establish entanglement connections with a predecessor node results in fewer qubits for the entanglement links with the successor.

Entanglement routing is a complex problem that involves not only establishing quantum links but also ensuring the reliable transmission of entanglement across a network. Qubit allocation and path selection has been considered as a pivotal part of entanglement routing in many previous studies such as \cite{shi2020concurrent, yang2022online, yang2023asynchronous} since establishing an entanglement link is probabilistic and unstable, relying on the availability of quantum channels between the parties and qubits on both ends, which directly influences the success of entanglement routing. However, these studies operate under the assumption of pre-known probabilities for creating entanglement links \cite{shi2020concurrent,yang2022online,zeng2022multi,zhao2021redundant,zhao2022segmented,zhao2022e2e,mao2023qubit,yang2023asynchronous,farahbakhsh2022opportunistic}. Meanwhile, they tend to overlook the critical aspect of addressing potential network attackers -- a well-explored area in conventional network security \cite{che2013routing,khan2016resource,zhou2019toward}. Such an attack can be performed on either the data qubit itself or the classical channel that delivers the Bell State Measurement result for quantum teleportation \cite{bouwmeester1997experimental}. Note that both categories of attackers can occur in any given time slot and be distinguished from regular entanglement connection failures, by no-cloning theorem or transmission errors in classical channels. 

In this paper, we focus on addressing the challenge of online optimal path selection and qubit allocation between two quantum nodes in the presence of a potential QDN attacker. We introduce a novel multi-armed bandits approach grounded, called adversarial group neural bandits. By developing an online adversarial optimization approach using multi-armed bandits, we provide a robust solution that optimizes both path selection and qubit allocation in real-time, without prior knowledge of success rates and under the presence of potential attackers. This enhances the overall efficiency and reliability of quantum entanglement routing, especially in dynamic and adversarial environments. Therefore, our approach contributes to advancing the broader field of entanglement routing by addressing a key challenge that has been largely overlooked in prior studies. Our key contributions encompass the following:
\begin{itemize}
    \item We present an online optimization problem that concurrently addresses path selection and qubit allocation between two quantum nodes. This scenario accounts for an attacker disrupting data qubit transmission and involves no prior knowledge of the success rate for creating each entanglement link. The goal is to maximize the long-term success rate for establishing long-distance entanglement connections between the chosen quantum nodes.
    \item To tackle the intricate nonlinear optimization objective, we frame the problem as an adversarial group neural bandits problem and propose a bandits algorithm, called EXPNeuralUCB, which treats possible paths as groups and performs qubit allocation as arm selection in each group. Moreover, by choosing suitable algorithm parameters, we theoretically prove that the algorithm has a regret upper bound of $O(T^{3/4}\sqrt{\log T})$, offering performance guarantees for our algorithm.
    \item We conduct comprehensive simulations to showcase the superiority of our proposed EXPNeuralUCB over other established advanced bandit algorithms.
\end{itemize}

\section{Related Work}

\textbf{Entanglement Routing.} Early research delved into specific network topologies like sphere, ring, diamond star, and chain configurations \cite{schoute2016shortcuts,pant2019routing,chakraborty2019distributed,vardoyan2019stochastic }. Recent studies have shifted focus towards a general QDN topology \cite{shi2020concurrent}, aiming to maximize expected network throughput while incorporating theoretical analyses for performance guarantees. An extension of this research \cite{zeng2022multi} addresses the challenge of efficiently utilizing network resources to support multiple SD (source-destination) pairs concurrently. Strategies have been explored to augment failure tolerance; some approaches leverage redundant entanglement links in routing to bolster the robustness of QDN \cite{zhao2021redundant, zhao2022segmented, zhao2022e2e}. Another avenue of exploration involves a qubit allocation algorithm in QDNs, which combines simulated-annealing and local search techniques \cite{mao2023qubit}. Additionally, an online entanglement routing scenario has been proposed \cite{yang2022online}, which involves processing requests upon arrival. As an extension, the entanglement routing paradigm in QDNs has evolved from the time slot mode to an asynchronous scheme \cite{yang2023asynchronous}, allowing more proactive utilization of idle quantum resources. Furthermore, opportunistic techniques have been introduced to QDNs \cite{farahbakhsh2022opportunistic}, enabling the opportunistic establishment of quantum links and augmenting routing flexibility. One work~\cite{wang2024adaptive} considers the qubit allocation and entanglement routing problem from a user-centric view with a long-term limited qubit budget.  However, none of these approaches address the lack of prior knowledge regarding the success rate of each entanglement link or consider adversarial scenarios.

\textbf{Multi-armed Bandits.} Bandit problems are typically categorized as either stochastic bandits or adversarial bandits based on how rewards are generated~~\cite{bubeck2012regret}. The classical UCB algorithm~\cite{auer2002finite} and EXP3 algorithm~\cite{auer2002nonstochastic} have been developed for optimal regret bounds in stochastic and adversarial bandits, respectively. However, existing approaches are limited to either the stochastic or adversarial regime, posing challenges for problems with coupled stochastic and adversarial rewards. Recent efforts address the intersection of stochastic and adversarial bandits. In one line of work~\cite{bubeck2012best,seldin2014one}, attempts are made to design a single algorithm applicable to both regimes without prior knowledge. Nevertheless, these works consider rewards from a single distribution. In another line of research~\cite{lykouris2018stochastic,gupta2019better}, the focus is on stochastic bandits with adversarial corruption of reward observations, though the actual received reward is unaffected. A recent study~\cite{huang2023adversarial} considers joint rewards influenced by both stochastic distribution and adversarial behavior. However, the stochastic part assumes linearity with the feature vector. The traditional approach to contextual bandits with general nonlinearity is by nonparametric models via Reproducing Kernel Hilbert Space (RKHS)~\cite{micchelli2006universal}, i.e, Kernel bandits~\cite{valko2013finite}. Thanks to the development of NTK theory~\cite{jacot2018neural},~\cite{zhou2020neural} first utilizes the NTK-based approximation on neural networks and presents a provably efficient MLP-based contextual bandit algorithm, i.e., NeuralUCB. Many follow-up works~\cite{ban2021ee, salgia2023provably} were inspired by NeuralUCB to model the non-linear reward function under the framework of NeuralUCB. However, these works do not consider the existence of an attacker.

\section{Background} \label{s2}

\begin{figure}
    \centering
    \includegraphics[width=0.4\textwidth]{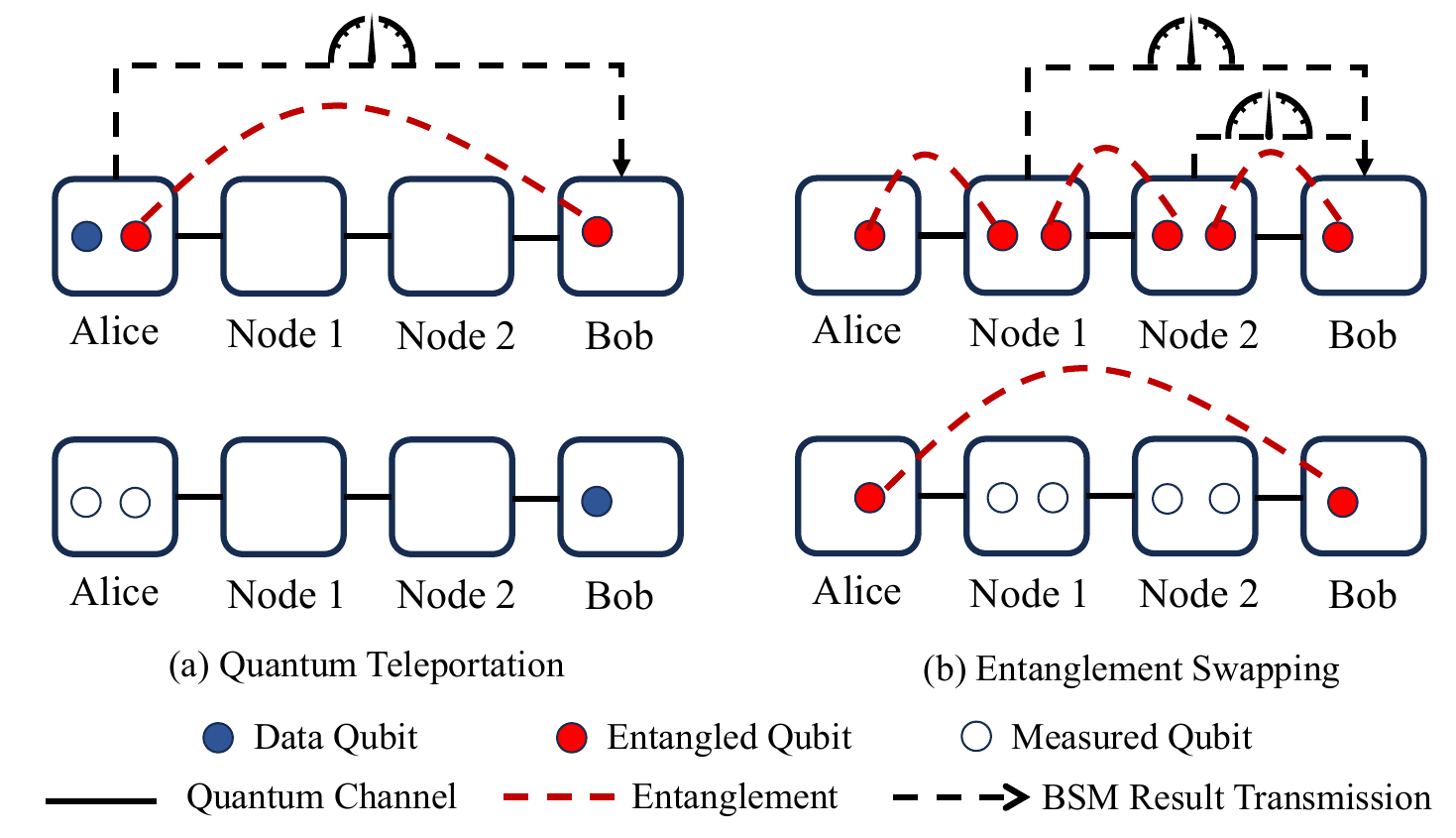}
    \vspace{-13pt}
    \caption{Quantum Teleportation and Entanglement Swapping.}
    \label{fig:quantum_opreation}
    \vspace{-15pt}
\end{figure}

\subsubsection{Quantum Teleportation}
Qubits, short for quantum bits, stand as the foundational unit of quantum information. Unlike classical bits restricted to 0 or 1 states, a qubit can simultaneously exist in a coherent superposition of these states. In a two-qubit system, a pair of entangled qubits is known as Bell pairs. One of the significant applications of this entanglement is quantum teleportation, depicted in Figure~\ref{fig:quantum_opreation}. When Alice and Bob share this pair of entangled qubits (referred to as entangled qubits), Alice can teleport the state of another qubit carrying data information (referred to as data qubits) to Bob. This complex procedure involves Alice conducting a Bell State Measurement between her data qubit and the shared entangled qubit, transmitting the measurement result to Bob through a classical channel. Upon receiving this information, Bob applies unitary operations to his entangled qubit. Consequently, Bob's entangled qubit replicates the state of the original data qubit, while Alice's data qubit collapses, disrupting the entanglement between the two entangled qubits.

\subsubsection{Entanglement Swapping}

In order to perform quantum teleportation between remote quantum nodes, establishing long-distance entanglement is the initial requirement. A crucial technique for achieving this is entanglement swapping, depicted in Figure~\ref{fig:quantum_opreation}. Here is an illustrative scenario: Alice shares an entangled qubit pair with a third party, Carol, while Bob also holds a qubit pair with Carol. Through a swapping operation on her qubits, Carol can teleport the state of her qubits, initially entangled with Alice, to Bob. As a result, Alice and Bob effectively possess an entangled qubit pair, despite lacking a direct connection. This facilitates the establishment of long-distance entanglement between distant parties. We assume a successful swapping operation due to recent advancements significantly enhancing its success rate to approximately 1, as also presumed in recent state-of-the-art studies \cite{yang2023asynchronous}. Even when considering the probability of swapping, incorporating a product term akin to the probability of entanglement links will not impact the efficacy of our algorithm.

\subsubsection{Quantum Data Network}
QDN is composed of quantum nodes and quantum channels, forming the nodes and edges of the network. Nodes connected via quantum channels are assumed to also be connected through classical channels. All quantum nodes can perform entanglement swapping and establish links with other nodes, but they are limited by their quantum memory capacity. Moreover, quantum channels are prone to losses, with the success rate of a single entanglement attempt dropping as low as $2.18 \times 10^{-4}$ \cite{stephenson2020high}. The decoherence time for entanglement is approximately 1.46 seconds, as noted in \cite{dahlberg2019link}, while each entanglement attempt takes about $165\mu$s \cite{stephenson2020high}. Consequently, only a limited number of thousands of entanglement attempts can be made on a single quantum link within a given time slot.

\section{System Model} \label{s3}
\subsection{Network Architecture}

We examine a QDN modeled as an undirected graph $\mathcal{G} = \langle\mathcal{V}, \mathcal{E}\rangle$, where $\mathcal{V}$ denotes the set of quantum nodes and $\mathcal{E}$ represents the set of edges. Each edge $e = (u,v) \in \mathcal{E}$ indicates the presence of quantum channels between the nodes $v$ and $u$. Every quantum node $v \in \mathcal{V}$ is equipped with a limited quantity of $Q_v$ qubits, and $W_e$ denotes the number of available quantum channels on edge $e$. Meanwhile, available qubits and channels may vary over time due to changes in the network circumstance, and we denote them as $Q_v^t$ and $W_e^t$ respectively if needed.

The establishment of a quantum link on an edge $e = (v, u)$ requires one qubit each from nodes $v$ and $u$ along with a quantum channel on $e$. However, successful entanglement establishment is not assured for every quantum channel. Let $\tilde{p}_e$ represent the unknown success probability of establishing a single entanglement link between nodes $v$ and $u$ during a single attempt. This probability is contingent on both the physical properties of the channel material and the distance between the two nodes. Typically, $\tilde{p}_e$ is low and can vary across different edges. In this study, we assume that the entanglement probability for each edge is unknown. 

To increase the probability of successful entanglement, nodes $v$ and $u$ can employ multiple quantum channels and make numerous attempts on each channel within a given time slot. Assuming the outcomes of these attempts are independent, the success probability on a single channel after $K$ attempts is given by $p_e = 1 - (1-\tilde{p}_e)^K$. Then the overall success probability using $q_e$ qubits at two ends of edge $e$, $u$ and $v$ respectively, is given by: $P_e(q_e) = 1 - (1 - p_e)^{q_e}.$ It is important to note that we have the constraint $q_e \leq \min (Q_v, Q_u, W_e)$.

\subsection{Problem Formulation}
Our primary focus revolves around tackling the challenge of entanglement path selection and qubit allocation. This involves establishing entanglement connections and maximizing the cumulative success probability over a specified duration of time slots $T$ between two chosen quantum nodes. Importantly, this is achieved without prior knowledge of the success rate of entanglement links between any two neighboring nodes.

Consider a set of potential paths between the source node and destination node, denoted as $\mathcal{R}$, and $|\mathcal{R}|=R$. These potential paths do not change over time since they are pre-determined based on the maximum qubit capacity and are independent of the traffic pattern. A path $r \in \mathcal{R}$ is defined as a subset of graph edges $\mathcal{E}$ that form a connected path between them. Given a specific path selection $r\in\mathcal{R}$ and qubit allocation $\mathcal{N}(r) = \{q_e(r), \forall e \in r\}$ for a path $r$, the entanglement success rate can be calculated as the product of the success probabilities of the individual edges on that path as follows:
\vspace{-5pt}
\begin{align}
h_r(\mathcal{N}(r)) = \prod_{e \in r} P_e(q_e(r))\label{eq:succ_func}
\end{align}
Here, $P_e(q_e(r))$ denotes the success probability of edge $e$ when $q_e(r)$ qubits are allocated at each node on both ends of edge $e$. The success rate of entanglement for each path is treated as an unknown function that requires learning. While our formulation represents this function as \eqref{eq:succ_func}, there might be additional factors we have not taken into account. This complexity and formulation challenge motivate us to employ a neural network for learning this function.

Moreover, we consider the potential presence of an adversary in the system, which aims to disrupt the quantum teleportation process and will select paths to attack during each time slot. This adversary may execute attacks either directly on the data qubits or on the classical channels transmitting the measurement result for quantum teleportation. The former attack is detectable due to the no-cloning theorem, while the latter can be identified through transmission errors in the classical channel. Both types of attacks fail the quantum teleportation operation and can be distinguished from failing to establish an entanglement connection in the normal case. Specifically, in each time slot $t$, we choose a path $r^t$ and perform qubit allocation $\mathcal{N}(r^t)$ to establish entanglement connections; the adversary simultaneously chooses a binary attack vector $a^t = (a^t(r), \forall r\in \mathcal{R})$, where $a^t (r) = 0$ if the adversary performs an attack on the path $r$ and $a^t(r)=1$ otherwise. At the end of time slot $t$, we can only observe the success or failure of establishing at least one entanglement connection, which can be denoted as a random variable $Y^t$. Note that this entanglement connection is typically verified by performing a Bell state measurement \cite{bouwmeester1997experimental} on the entangled qubits, which indirectly confirms the entangled state without directly measuring the qubits themselves. $Y^t$ conforms to the Bernoulli distribution, which takes the value $1$ with probability $s^t$ and $0$ with probability $1-s^t$, where $s^t$ is the unknown success rate depending on the selected path and the allocated qubits strategy and can be formulated as:
\begin{equation}\label{eq:succ_final}
    s^t(r^t, \mathcal{N}(r^t)) = a^t(r^t) \prod_{e \in r^t} P_e(q_e(r^t)).
\end{equation}
Note that when an attacked path is chosen, we can still differentiate the attack from a normal connection failure, as previously discussed, even though the latent success rate is 0, just like in the failure scenario. Clearly, we have $\mathbb{E}[Y^t] = s^t(r^t, \mathcal{N}(r^t))$.

\textbf{Objective}: The quantum user's objective is to maximize the success rate of entanglement connections between the source node and destination node over $T$ time slots ($t = 0, 1, ..., T-1$) by selecting an optimal path and allocating qubits along the path in each time slot:
\vspace{-10pt}
\begin{equation}
    \max \sum_{t=1}^T s^t(r^t,\mathcal{N}(r^t)).\label{obj_func}
\end{equation}

\section{Algorithm Design}
In this section, we first frame the problem of quantum entanglement path selection and qubit allocation in the presence of attackers as an adversarial group neural bandits problem. Subsequently, we introduce an algorithm named EXPNeuralUCB designed to address this specific problem. In Section V-D, we provide proof demonstrating that EXPNeuralUCB can attain a sublinear regret bound.

\subsection{Adversarial Group Neural Bandits}
We frame the selection of entanglement paths and qubit allocation as a sequential decision-making problem over $T$ rounds. We define $R$ groups of arms, denoted as $\mathcal{R} = \{1,2,\dots, R\}$, each corresponding to a set of potential paths. Within each group $r \in \mathcal{R}$, the arms represent various qubit allocation strategies, which differ across groups. Each group $r$ has arms of dimension $D_r$, corresponding to the number of links along path $r$. We denote these dimensions collectively by $\mathcal{D} = \{D_1, D_2, \dots, D_R\}$.

In each round $t$, for each group $r$, there are $|\mathcal{X}_r^t|$ available arms, where $\mathcal{X}_r^t \subseteq \mathbb{R}^{D_r}$ consists of $D_r$-dimensional vectors. Each vector $x \in \mathcal{X}_r^t$ represents a feasible qubit allocation strategy along path $r$, adhering to node qubit capacity constraints $Q^t_v$ for all nodes $v$, and link qubit channel capacity constraints $W^t_e$ for all links $e$. Each group $r$ is associated with a function $h_r$, defined on the domain $\mathcal{X}_r \subseteq \mathbb{R}^{D_r}$, representing the entanglement success rate on path $r$ in a stochastic environment, as described by equation~\eqref{eq:succ_func}. We obtain the available arm set by exhaustively exploring all possible combinations of qubit allocations along the nodes of the path, which guarantees that all feasible qubit allocation strategies within the capacity constraints are thoroughly evaluated.
This function does not account for potential adversarial actions and remains unknown to the learner.
The reason why we do not use $h_r(x)$ to represent equation~\eqref{eq:succ_final} directly is that NeuralUCB is under the stochastic assumption~\cite{zhou2020neural}. The reward for choosing an arm $x$ in group $r$ is given by $h_r(x)$, where $h_r$ is constrained such that $0 \leq h_r(x) \leq 1$ for any $x$ in any group $r$.

\textbf{Optimal Benchmark}.
Given that the adversary can adopt any attack strategy against the groups, we focus on the optimal benchmark among the group-static strategies in hindsight. A group-static strategy involves selecting a fixed group throughout the entire duration of $T$ rounds, while allowing the chosen arm to vary. For a given group $r$, the optimal arm from the set $\mathcal{X}r^t$ that maximizes the expected reward can be computed as $\xi_r^t \triangleq \arg\max{x \in \mathcal{X}_r^t} h_r(x)$, independent of the adversary's attacks. In the special case where the set of available arms $\mathcal{X}_r^t$ remains constant across all rounds, the optimal arm $\xi_r^t$ for group $r$ is also fixed, allowing the time index to be omitted.

With the optimal arms in each group in each round understood, the optimal group given an attacking sequence $a^1, ..., a^T$ is thus the one that maximizes the total reward, which corresponds to maximizing times of successful entanglement connection, or equivalently, maximizing the accumulative success rate of entanglement connections defined in  \eqref{obj_func},
\vspace{-5pt}
\begin{align}
    \gamma(a^1, ..., a^T) = \arg\max_{r \in \mathcal{R}} \sum_{t=1}^T s^t(r, \xi_r^t, a^t)
    \label{optimal}.
\end{align}
For notation simplicity, we write $\gamma(a^1, ..., a^T) = \gamma$ by dropping the attack sequences but the readers should be cautious that the optimal group $\gamma$ depends on the attacks (and $T$). 

\textbf{Regret}. To optimize the objective function specified in \eqref{obj_func}, we defined the regret of the learner as the difference between the total reward achieved by the optimal benchmark and the total reward attained by the learner's algorithm.
\vspace{-10pt}
\begin{align}\label{eq:reg_def}
    \reg(T) = \sum_{t=1}^T s^t(\gamma, \xi_\gamma^t) - \mathbb{E}[\sum_{t=1}^T s^t(r^t, x^t)],
\end{align}
where this expectation is over the possible internal randomization of the algorithm. The aim is to design a bandits algorithm that exhibits sublinear regret, indicating that the round-average regret diminishes to 0 as $T$ approaches infinity.

The bandit problem under consideration exhibits a semi-stochastic and semi-adversarial nature. On one hand, the task of learning the optimal arm within a group represents a stochastic non-linear bandit problem. On the other hand, learning the optimal group is framed as an adversarial bandit problem. Consequently, the problem at hand involves uncertainties arising from both the stochastic characteristics of the environment and the adversarial behavior of the opponent simultaneously. The learner's received reward is, therefore, a combined outcome influenced by both factors.
\subsection{Function Approximation via Neural Network}
We employ the NeuralUCB framework~\cite{zhou2020neural} to model the non-linear reward function. NeuralUCB utilizes neural networks for estimation in the following manner:

\textbf{Neural Network Architecture:} Let $f_r$ be  overparameterized multi-layer perceptions (MLPs)\footnote{For the simplicity of notation, we assume the width of each layer is $m$ and that MLPs for each group share the same set of hyperparameters, e.g., $m, L,\lambda,\zeta, J$ in Algorithm 1. In practice, their hyperparameters may vary from each other.} with depth $L \ge 2$ and width $m$ for each hidden layer to represent the unknown function $h_r$:
\begin{equation}
    f_r(x;\theta_r) = \sqrt{m}Z_r^L\sigma(Z^{L-1}_r\sigma(\dots\sigma(Z^1_rx))),
\end{equation}
where $\theta_r$ are stacked by $Z_r^{l} ,\forall l \in [L],\forall r \in \mathcal{R}$; Given $x \in \mathbb{R}^{D_r}$, $Z^1_r \in \mathbb{R}^{m \times {D_r}}$,$Z^l_r \in \mathbb{R}^{m\times m}$, $2 \le l \le L-1$,$Z^L_r \in \mathbb{R}^{m\times 1}$, $\sigma(x) = \max\{x,0\}$ is the rectified linear unit (ReLU) activation function. We denote the gradient of the neural network function by $g_r(x;\theta_r) = \nabla_{\theta_r} f(x;\theta_r)$.

For every $r \in \mathcal{R}$, the parameters $\theta_r$ follow similar initialization and update procedures as follows. 

\textbf{Neural Network Initialization:} We initialize $\theta_r$ with $\theta_{r}^0 = [{\text{vec}(Z^1_r)}^\top, \dots, {\text{vec}(Z^L_r)}^\top] \in \mathbb{R}^k$ with $k = m + mD_r + m^2(L-1)$, where for each $1 \le l \le L-1, Z_r^l = (Z, 0;0,Z)$, each entry of $Z$ is generated independently from $\mathcal{I}(0,4/m);Z_r^L = (z^\top,-z^\top)$, each entry of $z$ is generated independently from Gaussian $\mathcal{I}(0,2/m)$.

\textbf{Neural Network Update:} In round $t$, $\theta_r$ is updated to $\theta_{r}^t$, i.e., the optimal solution to the loss function for neural network training. The loss function is defined as 
    \begin{equation}\label{loss_func}
        \mathcal{L} (\theta) = \sum_{b=1}^t ( f_r(\{x^b_r\}; \theta) - h_r(x^b_r))^2/2 + m\lambda\|\theta - \theta^0_r\|_2^2/2,
    \end{equation}
where $\lambda$ is the regularization parameter. Set $\theta^{(0)} = \theta_r^0$ and we adopt gradient-based methods to optimize the loss function for $J$ steps with step size $\zeta$:
\begin{equation}\label{loss_update}
    \theta^{(j+1)} = \theta^{(j)}-\zeta\nabla\mathcal{L}(\theta^{(j)}), 
\end{equation}
where $j \in \{0,\dots,J-1\}$, then the estimation for $\theta_r$ at round $t$ is set $\theta_r^t = \theta^{(J)}$.

\subsection{EXPNeuralUCB}
EXPNeuralUCB combines strengths from both EXP3 and NeuralUCB. On one hand, it preserves an unbiased cumulative historical reward estimate, denoted as $S^t_r$, for each group, enhancing the process of group selection. On the other hand, EXPNeuralUCB maintains a parameter estimate ${\theta}_r^t$ for each group. These estimates are incrementally updated across rounds using equations \eqref{loss_func} and \eqref{loss_update}, facilitating arm selection within a group. The algorithm's pseudo-code is outlined in Algorithm \ref{algorithm1}, and we explain the algorithm's procedure below.

\textbf{Group Selection}: 
In each round $t$, EXPNeuralUCB calculates an unbiased estimate of the cumulative historical reward $S^{t-1}_r$ up to round $t-1$ for each group $r$. This estimation is based on past group and arm selections, as well as reward realizations, following \eqref{cum-reward}. Here, $\mathbf{1}\{\cdot\}$ denotes the indicator function, and $P^b$ represents the group sampling distribution calculated and utilized in round $b$. it is worth noting that we slightly abuse notation by using $P_e(q_e(r))$ to express the success probability of edge $e$ when $q_e(r)$ qubits are allocated to $e$ in the past. When computing $S^{t-1}_r$, the actual reward (i.e., whether the entanglement was successfully established) received in round $b$, denoted as $Y^b$, is added to the cumulative reward of group $r$ only if the selected group in round $b$ is $r$. This addition is followed by division by the selection probability. Using the updated $S^{t-1}_r$, a new group sampling distribution can be computed through \eqref{sampling}. This distribution is a weighted sum of two distributions with weights $1-\beta$ and $\beta$. The first distribution selects group $r$ proportionally to $\exp{(\eta S^{t-1}_r)}$, favoring groups with higher cumulative reward estimates and emphasizing exploitation. The second distribution is simply a uniform distribution, promoting the exploration of all groups with equal probability. The weights $1-\beta$ and $\beta$ adjust the trade-off between exploitation and exploration at the group level. With the group sampling distribution $P^t$, a group $r^t$ is then sampled and subsequently chosen in the current round.

\textbf{Arm Selection}: Subsequently, EXPNeuralUCB determines the best-estimated arm within the selected group $r^t$ based on \eqref{arm-estimate}. This computation utilizes the estimated group parameter ${\theta}^{t-1}_{r^t}$ and the auxiliary variable $V^{t-1}_{r^t}$. In \eqref{arm-estimate}, the first term represents the reward estimate of an arm $x$ in the group $r^t$, while the second term signifies the confidence associated with this estimate. The parameter $\alpha^t$ plays a crucial role in adjusting the balance between the exploitation and exploration of arms within each group as further illustrated in \eqref{alpha_def}.

\textbf{Variable Update}: Then EXPNeuralUCB updates the various variables depending on the present attacker strategy. Specifically, if the received reward $a^t(r^t) = 0$, which means the adversary attacks the selected group in round $t$, then all parameters are unchanged. Otherwise, for the selected group $r^t$, the auxiliary variable $V^t_{r^t}$ is first updated. Then update the group parameter  ${\theta}^t_{r^t}$ according to \eqref{loss_func} and \eqref{loss_update} to (approximately) minimize $\mathcal{L}(\theta)$ using gradient descent. For the unselected groups, their auxiliary variables and the parameter estimates also remain unchanged. 

\begin{algorithm}[t]
	\caption{EXPNeuralUCB}
	\begin{algorithmic}[1] 
		\State \textbf{Input}: Time horizon $T$, regularization parameter $\lambda$, $\theta^0_r \sim $ init(·), NeuralUCB exploration parameter $\nu$, confidence parameter $\delta$, step size $\zeta$, number of gradient descent steps $J$, network width $m$, network depth $L$, learning rate $\eta > 0$, EXP3 exploration rate $\beta \in (0, 1)$
        \State \textbf{Initialization}: $V^0_{r} = \lambda I_{{D}_{r}},\forall r \in \mathcal{R}$.
		\For {$t = 1 , ..., T$}
		    \State Compute estimated cumulative reward for each $r$
		        \begin{align}
		            S_{r}^{t-1} = \sum_{b = 1}^{t-1} \frac{\textbf{1}\{r^b = r\}}{P^b(r)}Y^b
		            \label{cum-reward}
		        \end{align}
		    \State Compute the sampling distribution for each $r$
        	    \begin{align}
        	        P^t(r) = (1-\beta)\frac{\exp(\eta S_{r}^{t-1})}{\sum_{r'=1}^R \exp(\eta  S_{r'}^{t-1})} + \frac{\beta}{R}
        	        \label{sampling}
        	    \end{align}		    
		    \State Sample group $r^t \sim P^t$
		    \State $\forall x \in \mathcal{X}_{r^t}^t$, compute $U_{r^t}^{t}(x)$ within $r^t$:
      		\begin{align}
		            U_{r^t}^{t}(x) =  f(x;\theta_{{r^t}}^{t-1}) + \alpha^t \|g_{r^t}(x;{\theta^{t-1}_{r^t}})/\sqrt{m}\|_{{(V_{r^t}^{t-1})}^{-1}}
		            \label{arm-estimate}
		        \end{align}
          \State Select the best estimated arm within $r^t$: $x^t = \arg\max_{x \in \mathcal{X}_{r^t}^t} U_{r^t}^{t}(x)$
		    \State Play group/arm $(r^t, x^t)$
		    \State Observe reward $r^t$
      \If{$a^t(r^t) = 0$}
		    \State Update $V_{r^t}^t = V_{r^t}^{t-1} + g_r(x^t;{\theta^{t-1}_{r^t}})g_r(x^t;{\theta^{t-1}_{r^t}})^\top/m $
		    \State Update ${\theta}_{r^t}^t$ by training MLPs $f_{r^t}$ using collected feedbacks as in \eqref{loss_func} and \eqref{loss_update}.
      \Else
      \State $V_{r^t}^t = V_{r^t}^{t-1}, \hat{\theta}_{r^t}^t = {\theta}_{r^t}^{t-1}$
      \EndIf
		    \State For all $r \neq r^t$, $V_{r}^t = V_{r}^{t-1}$, ${\theta}_{r}^t = {\theta}_{r}^{t-1}$
		\EndFor
	\end{algorithmic}
	\label{algorithm1}
\end{algorithm}

\subsection{Regret Analysis}
In EXPNeuralUCB, combating the stochastic non-linearity uncertainty within a group is intertwined with combating the adversarial uncertainty across groups. Thus, the regret analysis of EXPNeuralUCB must consider the regrets due to these two aspects simultaneously. In this subsection, we show that through a careful selection of the algorithm parameters, EXPNeuralUCB achieves a sublinear regret bound. 

We start with several lemmas on estimating the parameters of the groups.

\begin{lemma}[Lemma 5.1 in~\cite{zhou2020neural}] For a sufficiently large network width $m$, $\forall r \in \mathcal{R}, \forall x \in \mathcal{X}_r^t$, there exists a $\theta_r^*$ at round $t$ such that with probability at least $1-\delta$, we have
\begin{align}
    &h_r(x) = \langle g_r(x,\theta_r^0),\theta_r^*-\theta_r^0 \rangle \nonumber
    \\&\sqrt{m}\|\theta_r^* - \theta_r^0\|_2 \le \sqrt{2h_r^\top H_r^{-1}h_r},\nonumber
\end{align}
where $H_r$ is the neural tangent
kernel (NTK) matrix for $h_r$ defined in~\cite{zhou2020neural}. \end{lemma}

\begin{lemma} For a sufficiently large network width $m$, if 
\begin{align}
    &m \ge \text{poly}(T,L,\sup(\mathcal{D}),\lambda^{-1},\log(1/\delta)); \\& \lambda \ge \max\{1,{(2h^\top H^{-1}h)}^{-1}\}; \zeta = O\left({(mTL+m\lambda)}^{-1}\right),\nonumber
\end{align}
the estimated $\theta_r^t$ satisfies the following error bound for all group $r$ and round $t$ with probability at least $1-\delta$,
\begin{align}
    \|\theta_r^t - \theta_r^0\|_2 \leq 2&\sqrt{t/(m\lambda)} ,\|\theta_r^t - \theta_r^*\|_{V_r^t}
     \leq \alpha^t\sqrt{m},\nonumber
    \label{error_bound}
    \\ &\alpha^t = O(\sqrt{\tilde d\log(1+t)}),\nonumber
\end{align}
where $\tilde d$ is the effective dimension of the NTK matrix.
\end{lemma}

\begin{lemma}
    For a sufficiently large network width $m$, with probability of $1-\delta$, the single step regret bound for the group $r$ at $t$ round is bounded by $h_{r}(\xi_r) - h_{r}(x^t) \le 2\alpha^t\|g_{r}(x;{\theta^{t-1}_{r}})/\sqrt{m}\|_{{(V_{r}^{t-1})}^{-1}} + 3O(m^{-1/6})$.
\end{lemma}

\begin{lemma}
For a sufficiently large network width $m$, then we have for all $r$, 
\begin{align}
    &\sum_{t=1}^T \textbf{1}\{r^t = r\}a^t(r)\alpha^t \|g_r(x_r^t;\theta_r^{t-1}/\sqrt{m})\|_{{(V_r^{t-1})}^{-1}}\\
    \leq & O(\sqrt{\tilde d \log(1+T)})O(\sqrt{T\tilde d\log(1+T)}), \nonumber
\end{align}
with probability at least $1-\delta$. 
\end{lemma}

\begin{theorem}
For any $\delta \in (0, 1)$, by choosing $\beta = T^{-1/4}\sqrt{\log(T)}$ and $\eta = T^{-1/2}$, such that for any $\delta \in (0,1)$, if $m, \lambda, \zeta$ satisfy the same condition as Lemma 2, 
then with probability at least $1-\delta$, EXPNeuralUCB yields the following expected regret bound $\reg(T) = O(T^{3/4}\sqrt{\log(T)})$.
\end{theorem}

\textbf{Remark:} In the scenario where the optimal arm/allocation strategy in each group/path is known to the learner (pure-adversarial setting), the EXP3 algorithm yields a regret bound of $O(\sqrt{T})$. Conversely, in the pure-stochastic setting, where paths/groups are not attacked by the adversary in all rounds, NeuralUCB, specifically for the group-disjoint parameter case, achieves a regret bound of $\sqrt{T\log T}$. The regret incurred by EXPNeuralUCB is higher than both, attributable to simultaneously addressing adversarial uncertainty and stochastic uncertainty. However, it is noteworthy that neither EXP3 nor NeuralUCB can achieve a sublinear regret in our considered problem. Although EXPUCB\cite{huang2023adversarial} effectively addresses the challenge posed by the intersection of stochastic distribution and adversarial behavior, it faces limitations in learning the stochastic component with nonlinear features.

\section{Simulation Results}
In this section, we evaluate our proposed algorithm EXPNeuralUCB and compare its performance against several baselines. 

\subsection{Simulation Setup}
\subsubsection{Network setting} 
\begin{figure}
    \centering
    \includegraphics[width=0.4\textwidth]{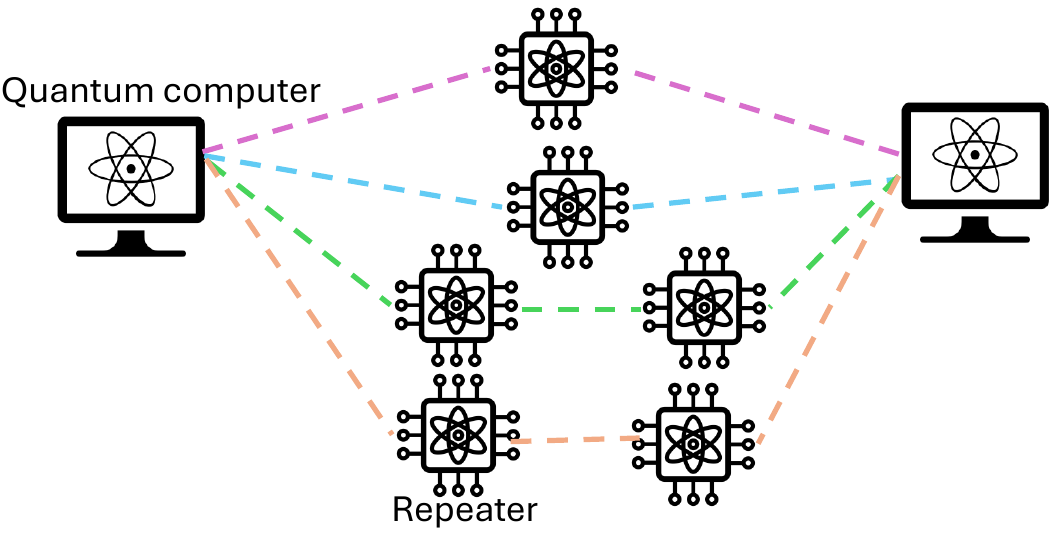}
\caption{{\it The topology of the QDN used in the simulation.}}
\label{fig:exp_set}
    \vspace{-15pt}
\end{figure}

We model a QDN with four potential paths ($R = 4$) connecting a source node to a destination node, as illustrated in Fig.~\ref{fig:exp_set}. The success probabilities for entanglement establishment $\tilde{p}_e$ vary across these paths and are detailed in Table~\ref{tab:exp}. To increase the probability of successful entanglement, each link undergoes 4000 entanglement connection attempts ($K = 4000$) in our simulations.

\subsubsection{Qubit capacity} We simulate a dynamic network environment within the QDN where the qubit availability at quantum repeaters fluctuates over time. This variation is influenced by the usage patterns of other network users. Specifically, we model two network states, ``busy'' and ``idle''. In the ``idle'' state, the qubit capacity at the quantum repeaters is greater than in the ``busy'' state. Additionally, the capacity of these repeaters varies across different paths, as detailed in Table~\ref{tab:exp}. It is assumed that both the source and destination quantum nodes possess a sufficient number of qubits, where "sufficient" means that their available qubits are equal to the maximum qubit capacity of any quantum repeater in the network. This ensures that the source and destination nodes can handle both "busy" and "idle" network states without being a bottleneck.

\begin{table}[htbp]
\vspace{-10pt}
\caption{Parameters of the Simulated QDN}
\vspace{-10pt}
\begin{center}
\begin{adjustbox}{width=0.9\columnwidth,center}
\begin{tabular}{|c|c|c|c|c|}
\hline
\textbf{Path}&\textbf{\# of Repeaters}&\textbf{Success Rate}&\textbf{Capacity (Busy/Idle)}\\
\hline
1 & 1 & $1.5e-4$ & 8/9 \\
\hline
2 & 1 & $1e-4$ & 10/11 \\
\hline
3 & 2 & $2e-4$ & 5/11, 5/11\\
\hline
4 & 2 & $1.5e-4$ & 6/12, 6/12\\
\hline
\end{tabular}
\label{tab:exp}
\end{adjustbox}
\end{center}
\end{table}

\subsection{Adversary's Strategy} 
We simulate two types of attacking strategies: an oblivious strategy and an adaptive strategy. In the oblivious strategy, the adversary employs a randomized Markov attacking strategy. The transition matrix for this strategy is:
\begin{equation}
\begin{bmatrix}
0.35 & 0.15 & 0.35 & 0.15 \\
0.3 & 0.2 & 0.3 & 0.2\\
0.35 & 0.15 & 0.35 & 0.15\\
0.3 & 0.2 & 0.3 & 0.2
\end{bmatrix},\nonumber
\end{equation}
where the entry in row $i$ and column $j$ denotes the probability of attacking path $j$ at time slot $t$ given that path $i$ was the target at time slot $t-1$. In the adaptive strategy, the adversary observes which path the learner chose in the previous time slot ($t-1$) and then targets the same path in the current slot ($t$). In both strategies, the adversary attacks exactly one path in each time slot.

\subsubsection{Baseline Schemes} We consider the following baselines in addition to the \textbf{Oracle} strategy in hindsight defined in equation~\eqref{optimal}.
\textbf{GNeuralUCB}: This variant adopts the classical NeuralUCB algorithm for the group setting, disregarding attacks. In each time slot, it selects the group and arm with the highest NeuralUCB of the estimated reward, according to equation~\eqref{arm-estimate}. The auxiliary variables remain unaltered if the received reward is 0. \textbf{EXPUCB}: This algorithm is similar to our proposed approach. It employs the EXP3 algorithm to choose the group based on the historical accumulated reward for each group. However, it leverages the LinUCB algorithm, as proposed by~\cite{huang2023adversarial}, for arm selection.

For EXPNeuralUCB, we use the Adam \cite{kingma2014adam} optimizer for neural network training
and set the default algorithm parameters as $\lambda =1,\delta = 0.1, m = 128, L=2, J=8, \zeta = 1\times10^{-4},\beta = T^{-1/4}\sqrt{\log T}, \nu = 1$ and $\eta = T^{-1/2}$.

\subsection{Performance Comparison}
\begin{figure*}[th]
	\centering
	\subfigure[Only busy period scenes]{
		\begin{minipage}[b]{0.3\textwidth}
			\includegraphics[width=0.95\textwidth]{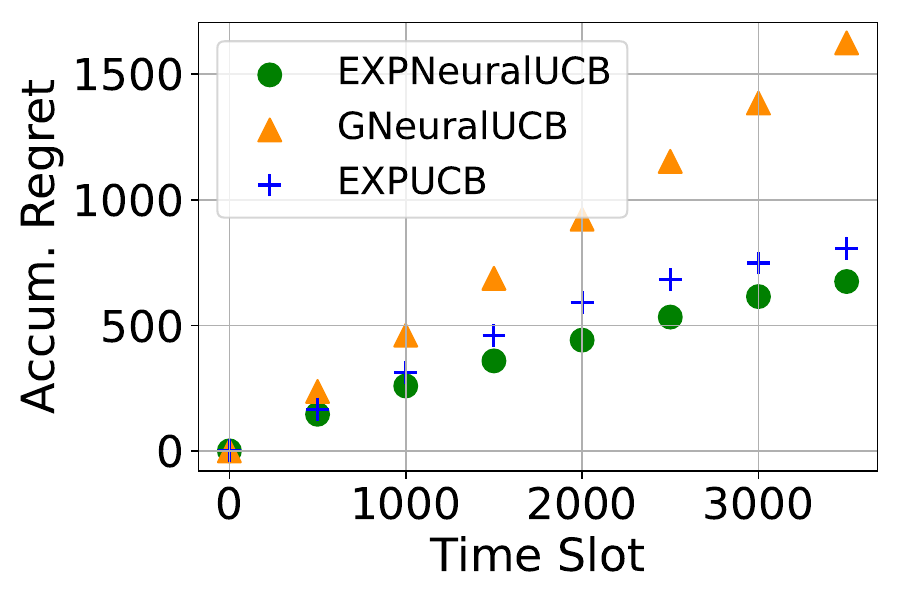}
		\end{minipage}
		\label{fig:regretYoLo}
	}
	\hspace{-6mm}
    	\subfigure[Only free period scenes]{
    		\begin{minipage}[b]{0.3\textwidth}
   		 	\includegraphics[width=0.95\textwidth]{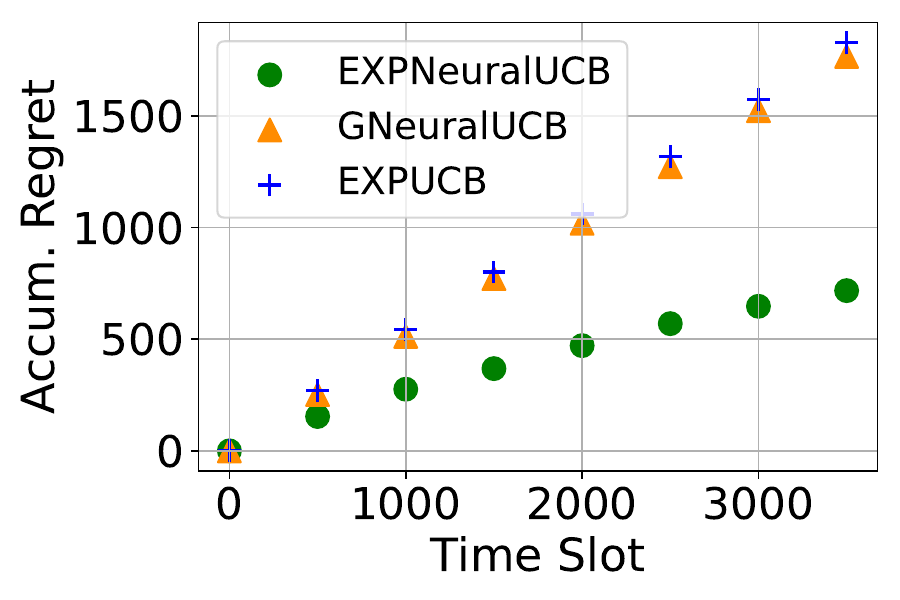}
    		\end{minipage}
		\label{fig:regretResNet}
    	}
    \hspace{-6mm}
    	\subfigure[50\% busy period 50\% free period]{
		    \begin{minipage}[b]{0.3\textwidth}
   	 	    \includegraphics[width=0.95\textwidth]{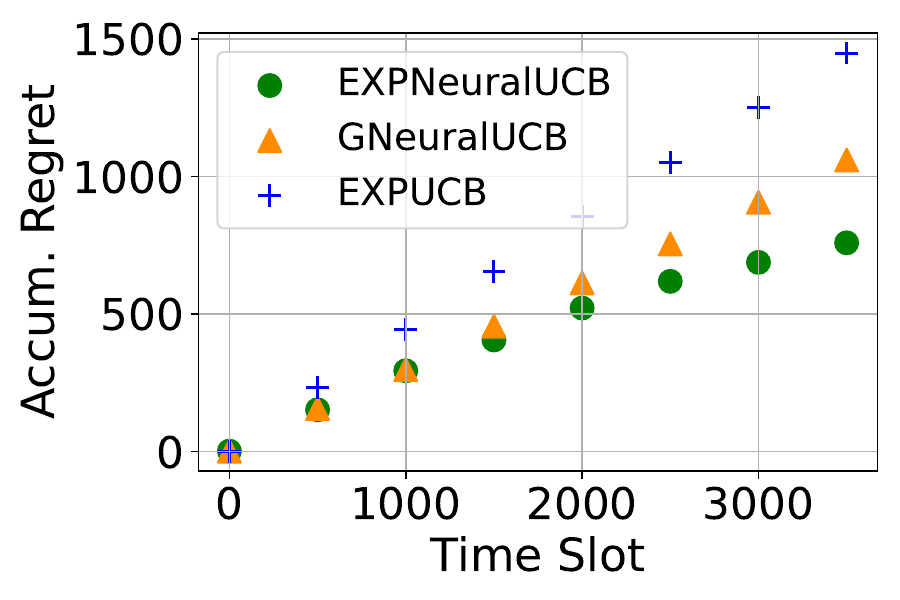}
		    \end{minipage}
	    \label{fig:regretMix}
	    }
	\caption{Regret achieved by EXPNeuralUCB and baselines.}
	\label{fig:regret}
\end{figure*}

\begin{figure*}[th]
	\centering
	\subfigure[Only busy period scenes]{
		\begin{minipage}[b]{0.3\textwidth}
			\includegraphics[width=0.95\textwidth]{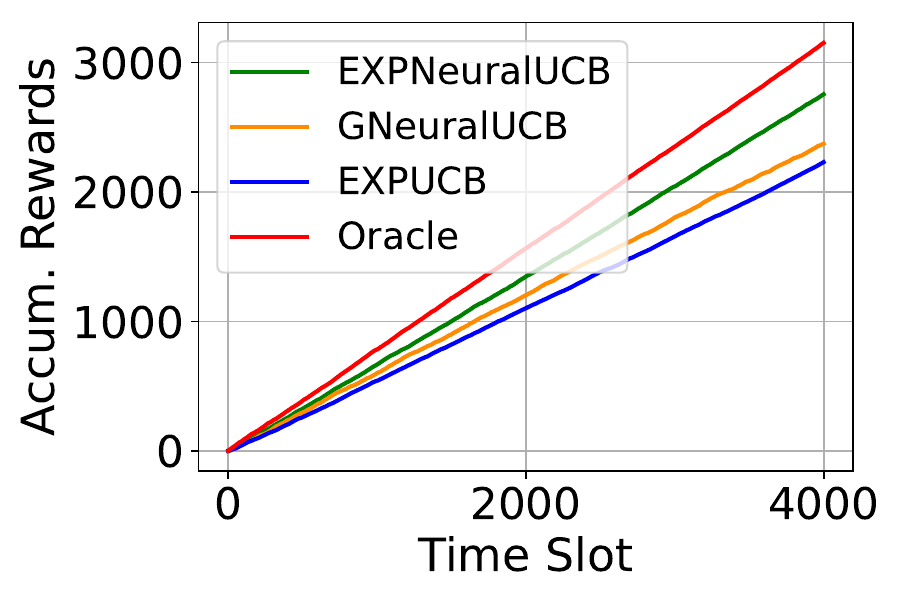}
		\end{minipage}
		\label{fig:rewardYoLo}
	}
	\hspace{-6mm}
    	\subfigure[Only free period scenes]{
    		\begin{minipage}[b]{0.3\textwidth}
   		 	\includegraphics[width=0.95\textwidth]{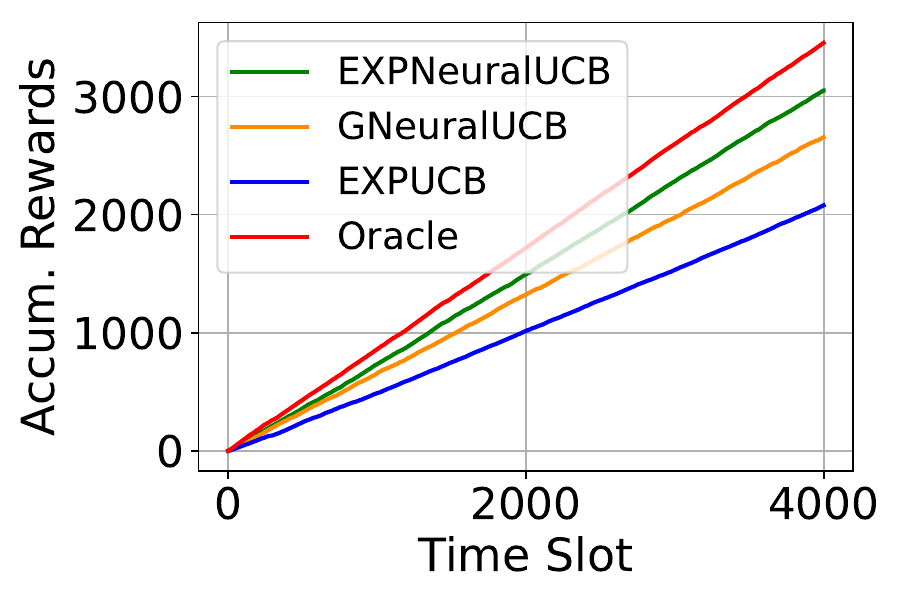}
    		\end{minipage}
		\label{fig:rewardResNet}
    	}
    \hspace{-6mm}
    	\subfigure[50\% busy period 50\% free period]{
		    \begin{minipage}[b]{0.3\textwidth}
   	 	    \includegraphics[width=0.95\textwidth]{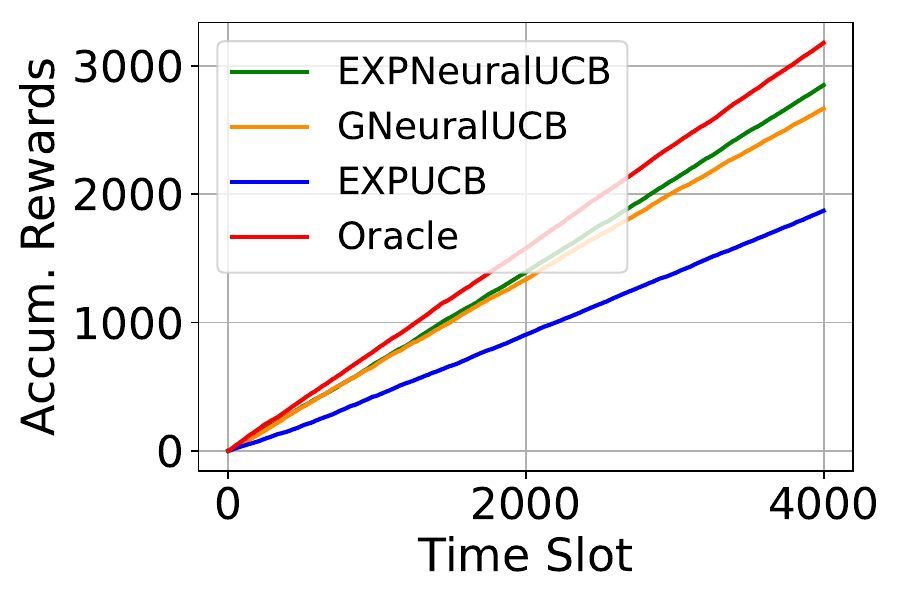}
		    \end{minipage}
	    \label{fig:rewardMix}
	    }
	\caption{Total reward achieved by EXPNeuralUCB and baselines.}
	\label{fig:reward}
\end{figure*}

We start by comparing the performance of EXPNeuralUCB with baseline algorithms in terms of total regret (as depicted in Fig.~\ref{fig:regret}) and total reward (illustrated in Fig.~\ref{fig:reward}), specifically under scenarios involving an oblivious attacker. The simulations cover three distinct network state distributions: ALL-BUSY, where every time slot is ``busy''; ALL-IDLE, where every slot is ``idle''; and HALF-HALF, where slots are ``busy'' or ``idle'' with equal probability of 0.5. It is noteworthy that the Oracle strategy may differ based on the number of slots, hence only selected data points are shown in Fig.~\ref{fig:regret}. The Oracle results presented in Fig.~\ref{fig:reward} apply specifically to simulations with $T = 4000$ slots. In all scenarios, EXPNeuralUCB outperforms non-Oracle baselines and exhibits sublinear regret throughout the simulation. The relative underperformance of non-Oracle baselines compared to EXPNeuralUCB can be attributed to specific limitations. \textbf{EXPUCB} adjusts to the attacking strategy and selects paths less vulnerable to attacks, yet it does not effectively learn the success probabilities of entanglement establishment across different quantum channels, thus failing to optimize qubit allocation. \textbf{GNeuralUCB}, on the other hand, overlooks the uncertainties introduced by path-level attacks, potentially choosing paths that suffer frequent attacks, resulting in lower overall rewards compared to EXPNeuralUCB.

\vspace{-15pt}

\subsection{Behaviors of EXPNeuralUCB}
\begin{figure*}[th]
	\centering
	\subfigure[Only busy period scenes]{
		\begin{minipage}[b]{0.3\textwidth}
			\includegraphics[width=0.95\textwidth]{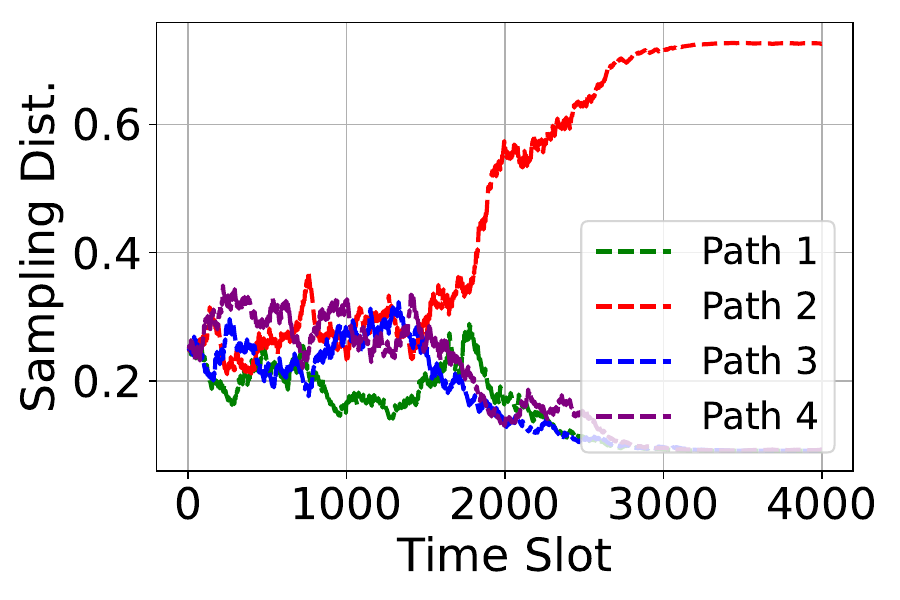}
		\end{minipage}
		\label{fig:samplingProbYoLo}
	}
	\hspace{-6mm}
    	\subfigure[Only free period scenes]{
    		\begin{minipage}[b]{0.3\textwidth}
   		 	\includegraphics[width=0.95\textwidth]{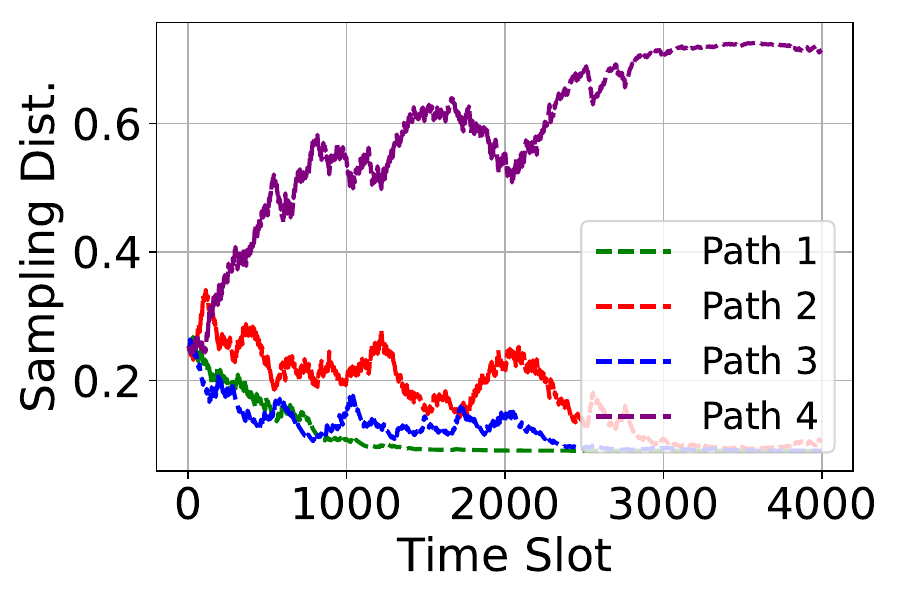}
    		\end{minipage}
		\label{fig:samplingProbResNet}
    	}
    \hspace{-6mm}
    	\subfigure[50\% busy period 50\% free period]{
		    \begin{minipage}[b]{0.3\textwidth}
   	 	    \includegraphics[width=0.95\textwidth]{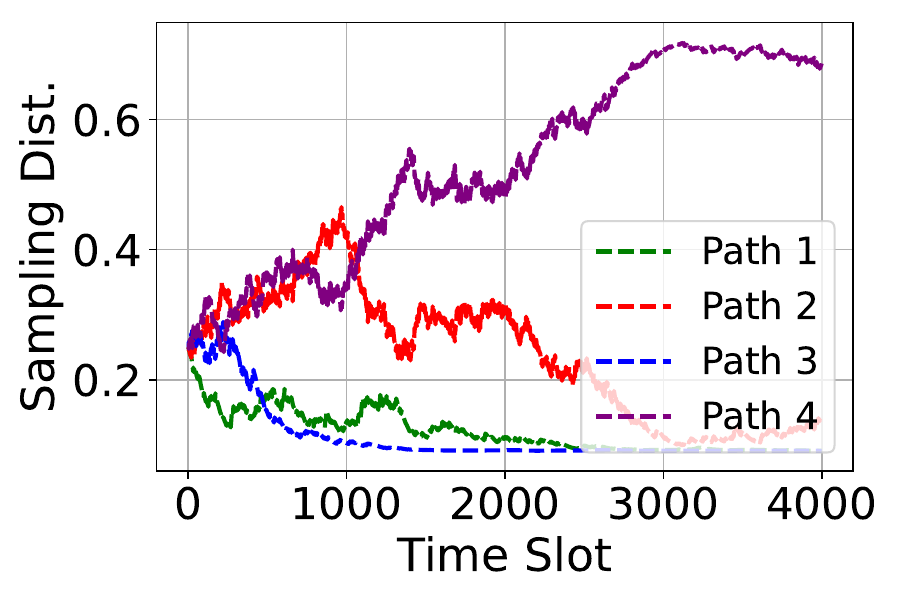}
		    \end{minipage}
	    \label{fig:samplingProbMix}
	    }
	\caption{Evolution of the sampling distribution for each path.}
	\label{samplingProb}
    \vspace{-10pt}
\end{figure*}

Let us explore in greater detail the performance of EXPNeuralUCB through an analysis of its path sampling distribution, depicted in Fig.~\ref{samplingProb}. Our simulation is configured to favor paths 1 and 2 in the ALL-BUSY scenario, with a preference for path 1, and paths 3 and 4 in the ALL-IDLE scenario, where path 3 is the optimal choice under no attack conditions. Fig.~\ref{fig:samplingProbYoLo} demonstrates that EXPNeuralUCB effectively identifies path 2 as the preferred choice in the ALL-BUSY scenario. Notably, it frequently avoids selecting path 1, even though it is the best option when there are no attacks, because it recognizes path 1's higher susceptibility to attacks. In contrast, Fig.~\ref{fig:samplingProbResNet} shows that in the ALL-IDLE scenario, EXPNeuralUCB primarily selects path 4, acknowledging that path 3—although optimal in an attack-free environment—is more prone to attacks. Furthermore, in the HALF-HALF scenario, where paths 3 and 4 offer higher overall rewards compared to paths 1 and 2 in the absence of attacks, EXPNeuralUCB, as illustrated in Figure \ref{fig:samplingProbMix}, accurately identifies path 4 as the best choice when under attack, consistent with the selections made by the Oracle strategy.

We also report the execution time and memory usage of EXPNeuralUCB and the compared algorithms, as shown in {Table~\ref{tab:time}}. The execution time and memory usage are averaged over 4000 rounds. As depicted in {Table~\ref{tab:time}}, EXPUCB has the smallest execution time and memory usage among the three algorithms, albeit with the worst performance. While EXPNeuralUCB and GNeuralUCB exhibit similar execution time and memory usage, EXPNeuralUCB outperforms GNeuralUCB in terms of overall performance. Hence, EXPUCB is suitable when minimizing execution time and memory is critical, whereas EXPNeuralUCB is a better choice for higher performance needs.
\vspace{-15pt}
\begin{table}[]
\caption{Execution Time and Memory Usage}
\vspace{-5pt}
\begin{adjustbox}{width=0.9\columnwidth,center}
\begin{tabular}{|c|c|c|c|}
\hline
\multicolumn{1}{|l|}{} & EXPNeuralUCB & GNeuralUCB & EXPUCB \\ \hline
Execution Time (s)     & 0.06275       & 0.06282     & 0.00161   \\ \hline
Memory Usage (MB)   & 334.03       & 333.30     & 314.30 \\ \hline
\end{tabular}
\label{tab:time}
\end{adjustbox}
\vspace{-15pt}
\end{table}

\subsection{Impact of Time-Varying State Distributions}
\begin{figure}
    \centering
    \includegraphics[width=0.48\textwidth]{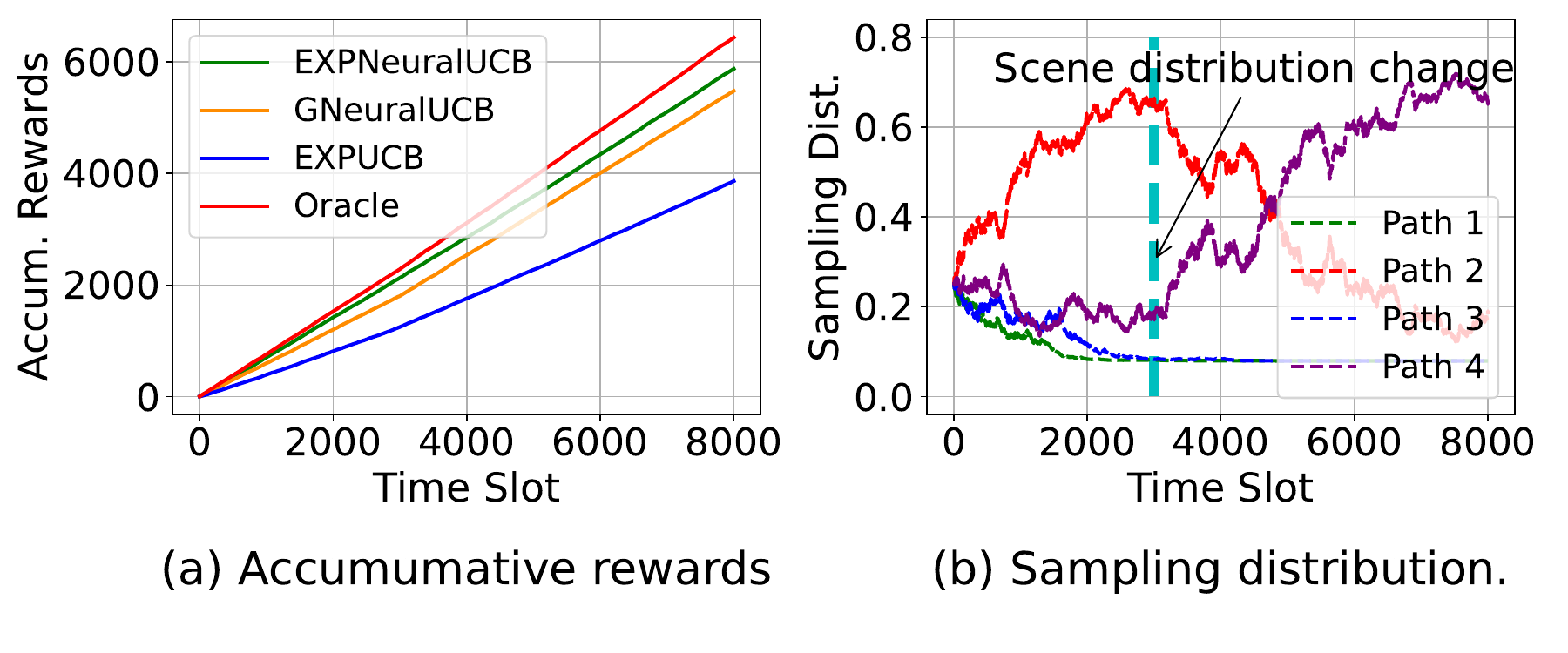}
 \vspace{-15pt}
\caption{{\it Performance of EXPNeuralUCB under changing period scenes.}}
\label{fig:dynamicArrivalTasks}
    \vspace{-15pt}
\end{figure}

In this set of experiments, we assess the performance of EXPNeuralUCB under conditions where the state distributions change over time. Initially, for the first 3000 slots, the probability of a slot being ``busy'' is 0.8 and ``idle'' is 0.2. From slot 3000 onwards, the probabilities invert, with ``busy'' at 0.2 and ``idle'' at 0.8, marking a significant shift in the state distribution at time slot 3000. Fig.~\ref{fig:dynamicArrivalTasks} illustrates the performance of EXPNeuralUCB in this dynamic environment. As depicted in Fig.~\ref{fig:dynamicArrivalTasks}(a), EXPNeuralUCB consistently achieves the highest total reward when compared to non-Oracle baseline strategies. This performance is attributed to its capacity to adapt and make strategic decisions in response to environmental changes. The adaptability of EXPNeuralUCB is further highlighted in Fig.~\ref{fig:dynamicArrivalTasks}(b), where there is a noticeable shift in the path sampling distribution following the change in state distribution. Initially, EXPNeuralUCB predominantly favors path 2; however, following the transition at slot 3000, it shifts to path 4.

\subsection{Impact of Attacker Strategies}
\textbf{Adaptive Attacking Strategy}. In this subsection, we examine the performance of EXPNeuralUCB when facing an adaptive attacking strategy, contrasting with the previously discussed oblivious attacking strategy. Notably, GNeuralUCB tends to perform poorly in this adaptive scenario due to its inability to adjust to ongoing attacks, often becoming stuck on paths that are consistently targeted by the adaptive strategy. To mitigate this, we have enhanced GNeuralUCB to create a new variant, NeuralUCB-Random. This modified version employs a strategy where the learner randomly selects a path each slot and then utilizes NeuralUCB to decide the allocation strategy.

\begin{figure}
    \centering
    \includegraphics[width=0.48\textwidth]{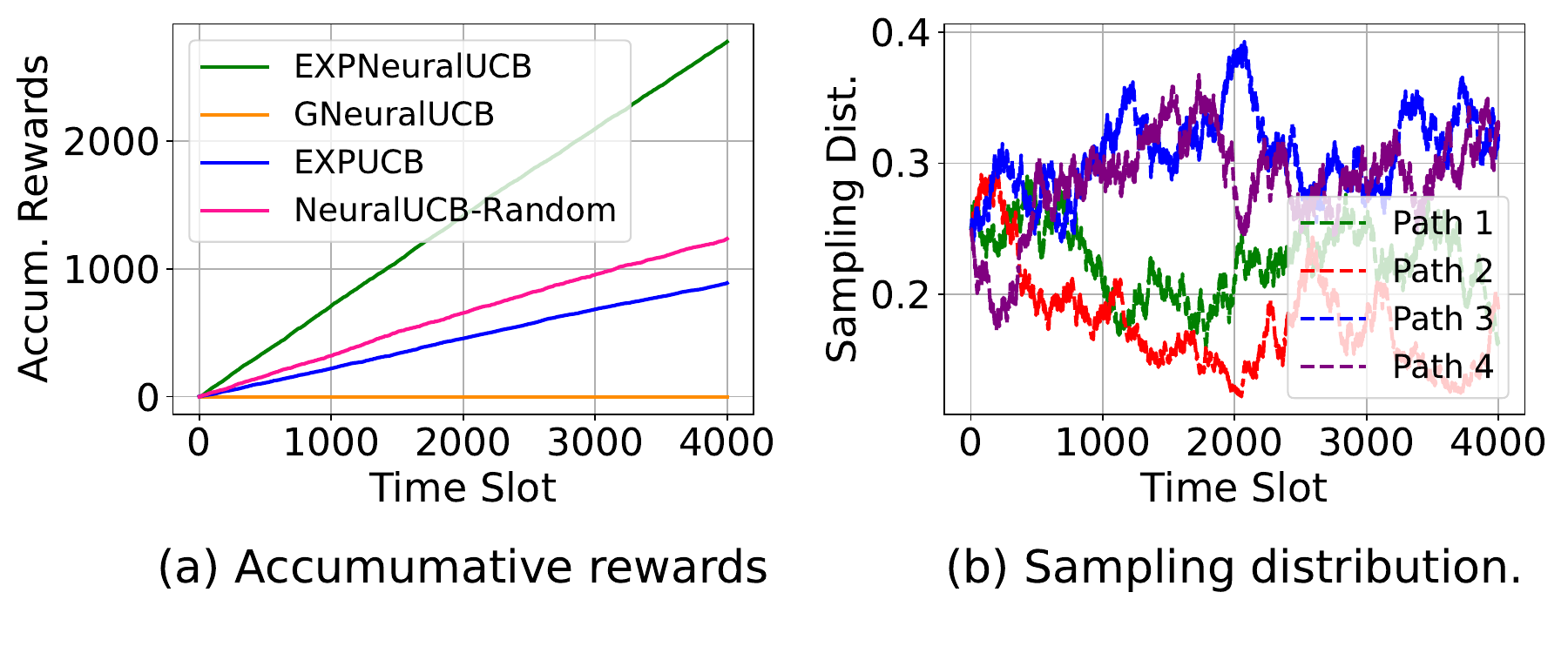}
 \vspace{-15pt}
\caption{{\it Performance of EXPNeuralUCB under the adaptive attacking strategy. (Only free period scenes).}}
\label{fig:AdaptiveAttacker}
    \vspace{-15pt}
\end{figure}

Fig.~\ref{fig:AdaptiveAttacker}(a) illustrates the total rewards achieved by EXPNeuralUCB and the baseline strategies in the ALL-IDLE scenario, where the discrepancy in performance is more pronounced than in scenarios with an oblivious attacker. This underscores EXPNeuralUCB's superior capability to adapt its path selection and allocation strategies in response to more complex and challenging conditions. Furthermore, Fig.~\ref{fig:AdaptiveAttacker}(b) sheds light on the path sampling probabilities over time. In the ALL-IDLE scenario, paths 3 and 4 are typically favored, and this preference is reflected in our simulation outcomes, where these paths are chosen more frequently. Remarkably, EXPNeuralUCB employs a strategic randomization between these paths, deviating from selecting the optimal path in an attack-free environment (path 3). This approach helps to circumvent persistent attacks linked with the adaptive attacking strategy. Although NeuralUCB-Random also incorporates path randomization, it does not match EXPNeuralUCB in identifying and selecting the most advantageous paths during attack-free periods, resulting in lower overall rewards.

\begin{figure}
    \centering
    \includegraphics[width=0.48\textwidth]{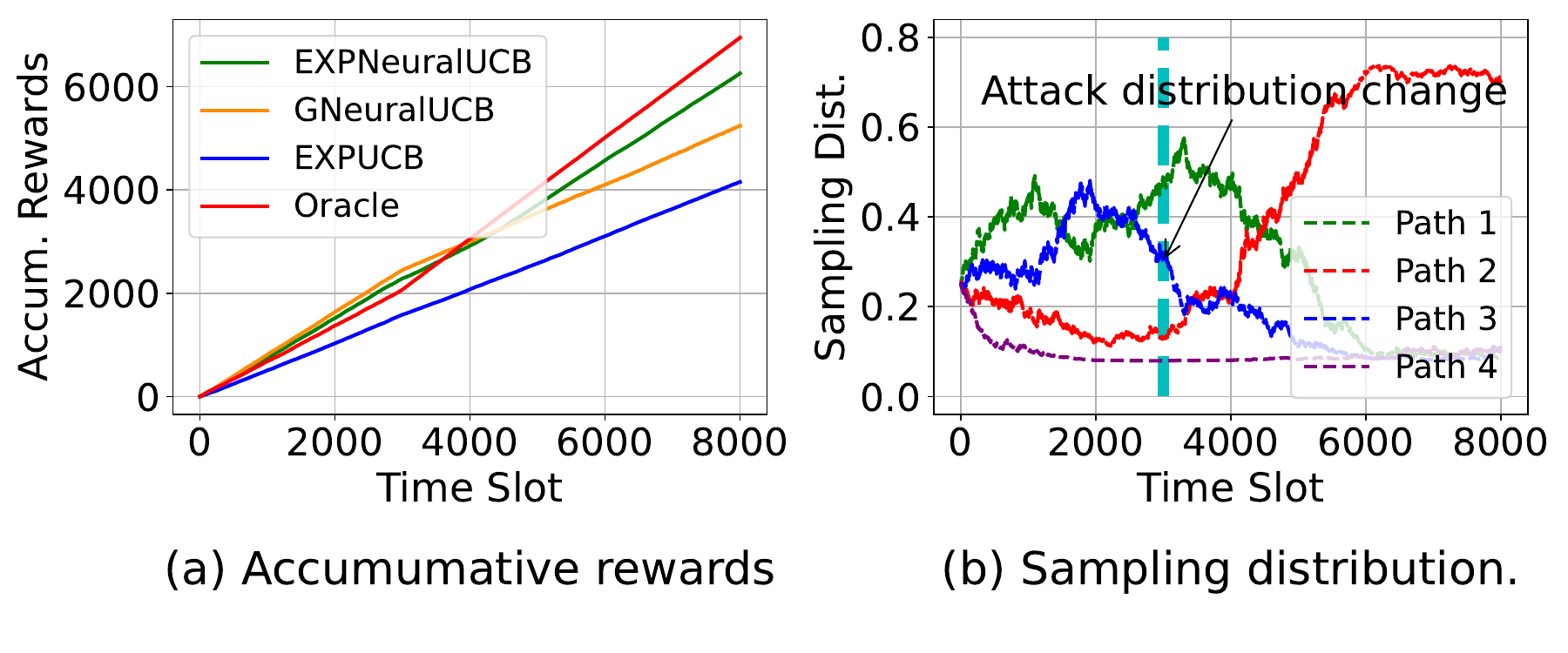}
 \vspace{-15pt}
\caption{{\it Performance of EXPNeuralUCB under the dynamic Markov attacking strategy (Only free period scenes).}}
\label{fig:dynamicMarkov}
    \vspace{-15pt}
\end{figure}

\textbf{Time-varying Attacking Strategy}. We conducted simulations to investigate how EXPNeuralUCB responds to dynamic strategies used by an oblivious attacker who alters the attacking transition matrix at time slot 3000. The matrix changes as follows:
\begin{equation}
\begin{bmatrix}
0.2 & 0.3 & 0.2 & 0.3 \\
0.15 & 0.35 & 0.15 & 0.35\\
0.2 & 0.3 & 0.2 & 0.3\\
0.15 & 0.35 & 0.15 & 0.35
\end{bmatrix} \Rightarrow
\begin{bmatrix}
0.35 & 0.15 & 0.45 & 0.05 \\
0.3 & 0.2 & 0.4 & 0.1\\
0.35 & 0.15 & 0.45 & 0.05\\
0.3 & 0.2 & 0.4 & 0.1
\end{bmatrix}.\nonumber
\end{equation}
Fig.~\ref{fig:dynamicMarkov}(a) shows the total rewards achieved by different algorithms across a span of 8000 time slots, with the Oracle providing a reference for optimal performance. In the initial 3000 slots, GNeuralUCB outperforms EXPNeuralUCB due to its selection of what is initially the optimal path (path 3), which  faces the least attacks. During this period, GNeuralUCB frequently selects this path more often than EXPNeuralUCB. However, after the attacking strategy shifts post-slot 3000 and path 3 ceases to be the most favorable, GNeuralUCB struggles to adapt. In contrast, EXPNeuralUCB demonstrates superior performance in adjusting to the new attack pattern, eventually surpassing GNeuralUCB in total reward. Fig.~\ref{fig:dynamicMarkov}(b) further illustrates the shift in path selection probabilities post-slot 3000, showcasing EXPNeuralUCB's ability to effectively adapt to changes in the attacker's strategy.

\vspace{-10 pt}

\section{Conclusion}
In this paper, we studied the problem of online path selection and qubit allocation in QDNs, specifically under the presence of potential path attacks. Our goal is to optimize the long-term success rate of entanglement connections between two quantum nodes. We approach this challenge by formulating it as an adversarial group neural bandits problem, introducing the EXPNeuralUCB algorithm, which treats potential paths as groups and qubit allocation as arm selections. Additionally, we demonstrate that EXPNeuralUCB achieves a theoretical regret upper bound of $O(T^{3/4}\sqrt{\log T})$. To assess the effectiveness of our algorithm, we conducted a series of experiments in various simulation environments, confirming that EXPNeuralUCB outperforms other baseline algorithms.

\section*{Appendix}
\subsection{Proof of Lemma 2}
\begin{proof}

This lemma follows Lemma 5.2 in \cite{zhou2020neural} by considering the sub-sequence of rounds in which the learner selects group $r$ and is not attacked by the adversary. Then we have : $\|\theta_r^t - \theta_r^*\|_{V_r^t} \le \alpha^t_r\sqrt{m}$
and $\alpha^t_r$ is defined as
\begin{align}\label{alpha_def}
    &\alpha^t_r =  \sqrt{1+C_1m^{-1/6}\sqrt{\log m}L^4t^{7/6}\lambda^{-7/6}} \times \\& \bigg(\nu\sqrt{\log\frac{\det V^t_{r}}{\det \lambda I} + C_2m^{-1/6}\sqrt{\log m} L^4t^{5/3}\lambda^{-1/6}-2\log\delta} \nonumber \\
    &  + \sqrt{\lambda}\bigg)\notag + (\lambda + C_3 tL)\Big[(1- \zeta m \lambda)^{J/2} \sqrt{t/\lambda} \nonumber \\&+ m^{-1/6}\sqrt{\log m}L^{7/2}t^{5/3}\lambda^{-5/3}(1+\sqrt{t/\lambda})\Big].\nonumber
\end{align}
for some constant $C_1,C_2,C_3$. $\lambda$ is the regularization parameter, $\nu$ denotes the exploration parameter for UCB-like estimation; $J$ is the number of gradient steps for each round of neural network training; when the network width $m$, regularization parameter $\lambda$, and step size $\zeta$ satisfy the same condition as Theorem 1,
$\alpha^t_r = O(\sqrt{\tilde d \log(1+T)})$ and the group index can be dropped.

\end{proof}
Note Lemma 2 leads to an upper confidence bound (UCB) on the prediction error of the estimated parameter $\theta_r^t$ for each group, and thus the UCB-based arm selection rule in \eqref{arm-estimate}. 

\subsection{Proof of Lemma 3}
\begin{proof}
    This Lemma follows the Lemma 5.3 in~\cite{zhou2020neural} by combining the Lemma 2.
\end{proof}

\subsection{Proof of Lemma 4}
\begin{proof}
Consider the sub-sequence of rounds where group $r$ is selected by the learner and not attacked by the adversary, we have
\vspace{-10pt}
\begin{align*}
    &\sum_{i=b}^{B} \sqrt{g_r(x^i;{\theta^{i-1}_{r}})^\top {(V_{r}^{i-1})}^{-1}g_r(x^i;{\theta^{i-1}_{r}})/m} \\
    \leq & \sqrt{B\sum_{b=1}^Bg_r(x^b;{\theta^{b-1}_{r}})^\top {(V_{r}^{b-1})}^{-1}g_r(x^b;{\theta^{b-1}_{r}})/m}
    \\ \leq & \sqrt{B\tilde d\log(1+B)} \leq  \sqrt{T\tilde d\log(1+T)}
\end{align*}
where the first inequality is due to Jensen's inequality, the second inequality is due to Lemma 5.4 in \cite{zhou2020neural}, and the last inequality is because the length of the sub-sequence $B$ is smaller than $T$. Further noticing that $\alpha^t  = O(\sqrt{\tilde d \log(1+T)})$ yields the desired bound.
\end{proof}

Now, we are ready to present the regret bound. 
\begin{proof}
Denote $w^t(r) = \exp(\eta S^{t-1}_r)$, $W^t = \sum_{r'=1}^R w^t(r')$ and $I^t(r) = \frac{\textbf{1}\{r^t = r\}}{P^t(r)}$. Then there exists a positive constant $\bar C$ such that for any $\delta \in (0,1)$, if $\eta$ and $m$ satisfy the same conditions as in Lemma 2, $W^{T+1}$ in the expectation can be lowered bounded with probability at least $1-\delta$ as follows
{\allowdisplaybreaks
\begin{align}
    &\mathbb{E}\left[\log\left(\frac{W^{T+1}}{W^1}\right)\right]\label{wt+1_wt}
    \geq \mathbb{E}\left[\log\left(\frac{\exp(\eta\sum_{t=1}^TI^t(\gamma) Y^t}{W^1}\right)\right] \\
    = & \mathbb{E}\left[\eta\sum_{t=1}^T  I^t(\gamma)Y^t\right] - \log R 
    = \eta\sum_{t=1}^T a^t(\gamma) h_{\gamma}(x^t) - \log R \nonumber\\
    \geq & \eta\sum_{t=1}^T a^t(\gamma) h_{\gamma}(\xi_\gamma) - \log R - \eta\sum_{t=1}^T a^t(\gamma) \times \nonumber\\
    & \left(2\alpha^t\|g_{\gamma}(x^t;\theta_{\gamma}^{t-1}/\sqrt{m})\|_{{(V_{\gamma}^{t-1})}^{-1}}
    +3O(m^{-1/6})\right) , \nonumber
\end{align}
}
where the first inequality uses $W^{T+1} \geq w^{T+1}(\gamma)$, the second equality uses $W^1 = R$, the third equality uses the definition of $Y^t$  and $\mathbb{E}[I^t(\gamma)] = 1$, and the last inequality is derived based on the Lemma 3. 

On the other hand, we have the following upper-bound
{\allowdisplaybreaks
\begin{align}
    &\mathbb{E}\left[\log\left(\frac{W^{t+1}}{W^t}\right)\right]
    =\mathbb{E}\left[\log\left(\sum_{r=1}^R \frac{\exp(\eta\sum_{b=1}^t I^b(r) Y^b)}{W^t}\right)\right] \nonumber\\
    =&\mathbb{E}\left[\log\left(\sum_{r=1}^R \frac{\exp(\eta\sum_{b=1}^{t-1} I^b(r) Y^b)}{W^t}\exp(\eta I^t(r) s^t)\right)\right] \nonumber\\
    =&\log\left(\sum_{r=1}^R \frac{P^t(r) - \frac{\beta}{R}}{1-\beta}\exp(\eta I^t(r) s^t)\right) \nonumber\\
    \leq & \log\left(\sum_{r=1}^R \frac{P^t(r) - \frac{\beta}{R}}{1-\beta}(1 + \eta I^t(r) s^t + (\eta I^t(r) s^t)^2)\right) \nonumber\\
    \leq &\sum_{r=1}^R \frac{P^t(r) - \frac{\beta}{R}}{1-\beta}(1 + \eta I^t(r) s^t + (\eta I^t(r) s^t)^2) - 1 \nonumber\\
    \leq & \sum_{r=1}^R \frac{P^t(r)}{1-\beta}(\eta I^t(r) s^t + (\eta I^t(r) s^t)^2) - \frac{\eta \beta}{R(1-\beta)}\sum_{r=1}^R I^t(r) s^t, \nonumber
\end{align}
}
where the first equality uses the definition of $W^{t+1}$, the second equality breaks the sum into two parts and uses $\mathbb{E}[Y^b] = s^b$, the third equality uses the definition of the sampling distribution $P^t$, the fourth inequality uses $e^z \leq 1 + z + z^2, \forall z \leq 1$, the fifth inequality uses $\log z \leq z -1, \forall z \geq 0$, and the last inequality holds by canceling out terms and realizing that $-\sum_{r=1}^R(\eta I^t(r) s^t)^2 \leq 0$. 
Noticing that $\sum_{t=1}^T \log\frac{W^{t+1}}{W^t} = \log \frac{W^{T+1}}{W^1}$, we can sum both sides for $t = 1, ..., T$ and compare with the lower bound in \eqref{wt+1_wt} and obtain
{\allowdisplaybreaks
\begin{align}
    &\eta\sum_{t=1}^T a^t(\gamma)(h_{\gamma}(\xi_\gamma)) - \log R - \eta\sum_{t=1}^Ta^t(\gamma)\times\label{confidence_upper_bound}\\
    &  \left(2\alpha^t\|g_{\gamma}(x^t;\theta_{\gamma}^{t-1}/\sqrt{m})\|_{{(V_{\gamma}^{t-1})}^{-1}} +3O(m^{-1/6})\right) \nonumber\\
    \leq & \sum_{t=1}^T (\sum_{r=1}^R \frac{P^t(r)}{1-\beta}(\eta I^t(r) s^t + (\eta I^t(r) s^t)^2) \nonumber\\
    &- \frac{\eta \beta}{R(1-\beta)}\sum_{r=1}^R I^t(r) s^t). \nonumber
\end{align}}
Reordering and multiplying both sides by $\frac{1-\beta}{\eta}$ gives
\vspace{-10pt}
\begin{align}
    &\sum_{t=1}^T \left( a^t(\gamma) h_\gamma(\xi_\gamma) - \sum_{r=1}^R\textbf{1}\{r^t = r\} s^t\right)\\
    \leq & \frac{1-\beta}{\eta}\log R + \sum_{t=1}^T\sum_{r=1}^R \eta I_t(r) (s^t)^2  \nonumber\\
    & + \beta \sum_{t=1}^T \left( a^t(\gamma)h_\gamma(\xi_\gamma) - \frac{1}{R} \sum_{r=1}^R I^t(r) s^t\right) + (1-\beta) \times \nonumber\\
    & \sum_{t=1}^T a^t(\gamma)  \left(2\alpha^t\|r_{\gamma}(x^t;\theta_{\gamma}^{t-1}/\sqrt{m})\|_{{(V_{\gamma}^{t-1})}^{-1}} +3O(m^{-1/6})\right) \nonumber
\end{align}
Now, consider the definition of the regret in \eqref{eq:reg_def},

\begin{align}
    &\reg(T) = \sum_{t=1}^T \left( s^t(\gamma, \xi_\gamma^t) - \sum_{r=1}^R\textbf{1}\{r^t = r\} s^t\right)\\
    \leq &\frac{1-\beta}{\eta}\log R + \eta RT + \beta T 
    + \frac{1-\beta}{\beta} \sum_{t=1}^T a^t(\gamma)\times \nonumber \\&\left(2\alpha^t\|r_{\gamma}(x^t;\theta_{\gamma}^{t-1}/\sqrt{m})\|_{{(V_{\gamma}^{t-1})}^{-1}} +3O(m^{-1/6})\right)\nonumber\\
    \leq&\frac{1}{\eta}\log R + \eta RT + \beta T +\frac{1-\beta}{\beta} \times \nonumber\\
    &\left[O(\sqrt{\tilde d \log(1+T)})O(\sqrt{T\tilde d\log(1+T)}) + 3O(m^{-1/6})\right], \nonumber
\end{align}
where the last inequality holds for Lemma 4 and sufficiently large $m$.

Finally, by setting $\beta = T^{-1/4}\sqrt{\log(T)}$ and $\eta = T^{-1/2}$, we have $\reg(T) = O(T^{3/4}\sqrt{\log(T)})$.
\end{proof} 

\bibliographystyle{IEEEtran}
\bibliography{IEEEabrv,main.bib}

\end{document}